\documentclass[aps,prl, twocolumn]{revtex4-2}
\usepackage{indentfirst}
\usepackage{color}
\usepackage{graphicx}
\usepackage{epstopdf}
\usepackage{subfigure}
\usepackage{float}
\usepackage{bm}
\usepackage{dcolumn}
\usepackage{hyperref}
\usepackage{amsmath}
\usepackage{comment}
\usepackage[sort]{cleveref}
\usepackage{lineno}
\begin{document}
\title{Interference-induced entanglement in an effectively zero-lifetime particle pair}
\author{Xin Wu, Xinbai Li}
\author{Zebo Tang}\email[Corresponding author, Zebo Tang Address: No. 96 Jinzhai Road, Hefei city, ]{Email: zbtang@ustc.edu.cn}
\author{Yusong Wang}
\author{Wangmei Zha}\email[Corresponding author, Wangmei Zha Address: No. 96 Jinzhai Road, Hefei city, ]{Email: first@ustc.edu.cn}

\affiliation{State Key Laboratory of Particle Detection and Electronics, University of Science and Technology of China, Hefei 230026, China and Department of Modern Physics, University of Science and Technology of China, Hefei 230026, China}

\begin{abstract}

Quantum entanglement in high-energy collisions is often obscured by finite lifetimes, dynamical evolution, and final-state interactions, complicating the identification of genuinely quantum correlations. Ultra-peripheral heavy-ion collisions provide a clean benchmark via the Drell–Söding production of nonresonant pion pair, realizing an effectively zero-lifetime particle pair whose quantum correlations are fixed at production and remain robust against subsequent elastic scattering. The coherent superposition of photoproduction amplitudes from two indistinguishable nuclei encodes the linear polarization of quasi-real photons in the orbital motion of the pair, generating a nonfactorizable two-particle quantum state. This entanglement leaves a direct experimental imprint: a characteristic second-harmonic azimuthal modulation in momentum space arising from spin-dependent interference between the two sources. In this paper, we establish a quantitative framework for Drell–Söding pion-pair production in relativistic heavy-ion collisions and predict the magnitude and transverse-momentum dependence of the entanglement-induced azimuthal asymmetry. Our results provide experimentally accessible signatures of interference-induced entanglement and a controlled test of quantum coherence in relativistic environments.
\end{abstract}

\maketitle

Quantum entanglement has been prepared and verified with high precision in controlled quantum optical systems, such as those involving cold atoms, where photon-mediated interactions generate and probe entangled states under well-isolated conditions~\cite{Bouwmeester:1997slj,Szigeti:2020uym,Cooper:2022onc}. A complementary frontier is to ask how entanglement behaves in open, rapidly evolving relativistic systems, where it is not engineered by design but can emerge from the coherent superposition of complex scattering amplitudes—providing a direct arena to probe the quantum-to-classical transition and the mechanisms of decoherence. High-energy particle and nuclear collisions offer access to such extreme conditions~\cite{Chen:2024aom} and have motivated growing interest in entanglement-related observables~\cite{Tu:2019ouv,Afik:2020onf,Gong:2021bcp,Barr:2024djo}. In particular, linearly polarized photons~\cite{Krauss:1997vr,Li:2019yzy,Wu:2022exl} in ultra-peripheral heavy-ion collisions (UPCs) have been shown to induce characteristic angular modulations~\cite{Xing:2020hwh,Zha:2020cst} that reflect the underlying coherence and interference of the production amplitudes~\cite{STAR:2022wfe,Brandenburg:2024ksp}，leading to a nonseparable two-particle state whose entanglement is encoded in the measured azimuthal asymmetry. A key open issue, however, is whether and how these observed correlations can be quantitatively tied to coherence fixed at the moment of production, rather than being reshaped by subsequent dynamical evolution and interactions~\cite{Baker:2017wtt}? 

Finite lifetimes can, in principle, alter relative phases or redistribute correlations among different degrees of freedom, complicating the connection between experimentally accessible observables and their production-stage origin. For photoproduction through a resonant intermediate state such as the $\rho^0$ meson, the finite lifetime introduces an intrinsic time window during which different production amplitudes continue to evolve. Relative phases associated with spin or polarization interference can accumulate during this interval and are subsequently averaged over the uncontrolled decay time of the resonance, leading to a suppression of the interference terms in experimentally accessible observables. In addition, the propagation and decay of the resonant state can encode partial which-source information into unobserved degrees of freedom, further reducing the visibility of quantum interference. As a result, the observed correlations in the final state may no longer faithfully reflect the quantum coherence fixed at the production stage.

In this context, the zero-lifetime limit plays a special conceptual role, serving as a clean benchmark in which no intermediate time evolution can modify quantum correlations. UPCs naturally realize such a scenario through the Drell-S${\rm\ddot{o}ding}$ (DS) mechanism~\cite{Drell:1960zz,Soding:1965nh}. In this process, nonresonant particle pairs are produced via the coherent superposition of photoproduction amplitudes associated with the two colliding nuclei~\cite{PhysRevLett.84.2330,Zha:2018jin}, which act as spatially separated yet indistinguishable photon sources. The observed smooth, non-resonant continuum implies an infinite uncertainty in the invariant mass (or width) of the production process. By the energy-time uncertainty principle, this infinite width directly corresponds to an effective lifetime of zero for any transient intermediate state, akin to a virtual exchange. Therefore, in this physical picture, the pair can be viewed as originating from the instantaneous decay of an effectively zero-lifetime vector meson-like state. The orbital motion of the pair is imprinted with the system’s initial angular momentum via the linear polarization of the quasireal photons, thereby correlating the two particles into a nonseparable quantum state.  While elastic scattering during the subsequent evolution may modify the overall momentum distribution, it does not affect the spin- and polarization-dependent interference observables considered here.
\begin{figure*}[t]
    \centering
    \includegraphics[width=0.9\linewidth]{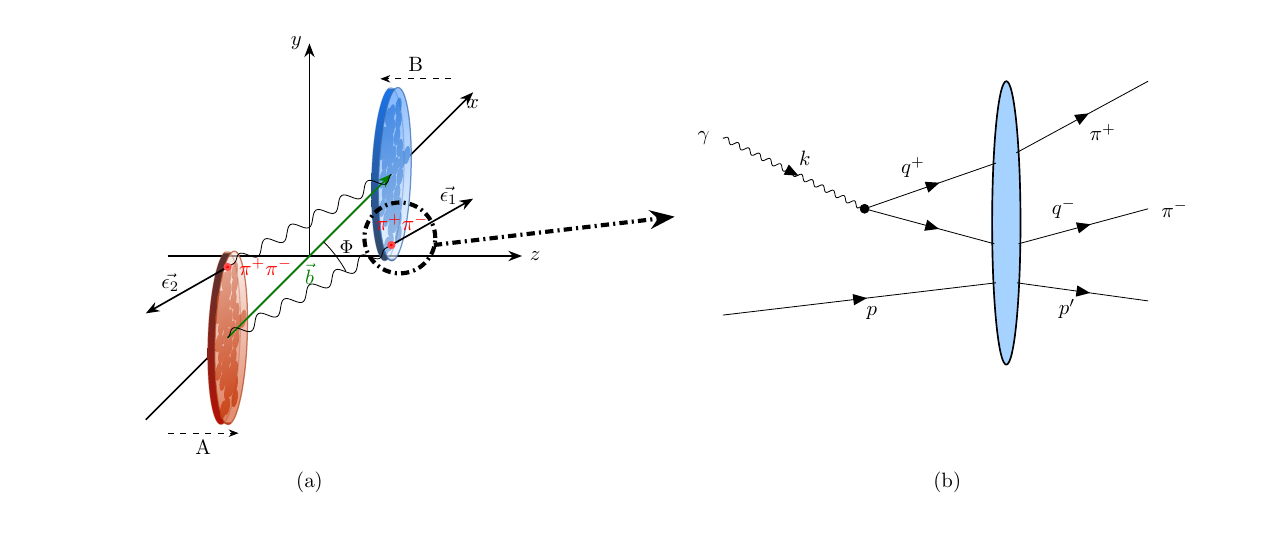}
    \caption{(a) Schematic illustration of the coordinate system used in this paper. The definition of the $x$, $y$, and $z$ axes is described in the text. $\epsilon$ is the polarization vector. $\Phi$ is the azimuthal angle between the polarization vector and the impact parameter. (b) Feynman diagrams for the Drell-S${\rm\ddot{o}ding}$ mechanism. The incident photon fluctuates into a virtual $\pi^+\pi^-$ pair, which subsequently undergoes elastic scattering off the nuclear target. The photon momentum is denoted by $k$, and the $\pi^\pm$ momenta by $q^\pm$. The proton momentum before (after) scattering is $p$ ($p'$).}
    \label{Fynmandiagram}
\end{figure*}

The DS mechanism has long been studied in photon-hadron reactions both experimentally and theoretically~\cite{Bingham:1973fu,Barber:1979ah,Bonn-CERN-EcolePoly-Glasgow-Lancaster-Manchester-Orsay-Paris-Rutherford-Sheffield:1980gri,Bodenkamp:1984dg,ZEUS:1995bfs,Afzal:2025nfe,Pumplin:1970kp,Ryskin:1997zz,Lebiedowicz:2014bea}, provides a well-established baseline for nonresonant particle-pair production. Ultra-peripheral heavy-ion collisions, however, introduce a qualitatively new two-source environment,  in which coherent photon fields from both nuclei contribute simultaneously to the production process. This additional geometric coherence introduces interference effects that are absent in single-source reactions and are not captured by the conventional DS framework, motivating its extension to the UPC case explored in this work.

Furthermore, the relatively broad width of the $\rho^0$ resonance allows the nonresonant $\pi^+\pi^-$ continuum to be cleanly separated from the resonant contribution over a wide invariant-mass range. In addition, coherent $\rho^0$ photoproduction in UPCs has been extensively measured~\cite{STAR:2007elq,ALICE:2015nbw,STAR:2017enh}, providing quantitative constraints on the resonant amplitude. These features make the $\pi^+\pi^-$ channel a controlled setting for studying the interference structure associated with the DS process.

To quantitatively describe this interference structure and its observable consequences, we formulate the production amplitude in a framework that explicitly incorporates the photoproduction geometry and photon polarization. We first adopt a right-handed coordinate system (Fig.~\ref{Fynmandiagram}(a)) defined in the photon-nucleon center-of-mass frame, with the $z$ axis along the momentum of the $\pi^+\pi^-$ pair (in experiments, $z$ is approximated by the incoming beam direction), the $y$ axis normal to the photoproduction plane, and the $x$ axis completing the orthonormal basis. The transverse spatial distribution of the DS photoproduction amplitude can then be written as the convolution of the photon flux amplitude and the corresponding $\gamma A \rightarrow \pi^{+}\pi^{-}A$ scattering amplitude: 
\begin{equation}
A\left(\vec{x}_{\perp},M_{\pi^{+}\pi^{-}}\right)=\left|\vec{a}\left(\omega,\vec{x}_{\perp}\right)\right|\Gamma_{\gamma A \rightarrow \pi^{+}\pi^{-}A}\left(M_{\pi^{+}\pi^{-}},\vec{x}_{\perp}\right)
\end{equation}
where $a\left(\omega,\vec{x}_{\perp}\right)$ is the photon flux amplitude generated by the heavy nuclei which can be given by the equivalent photon approximation (EPA)~\cite{Krauss:1997vr}:

    \begin{gather}
	\vec{a}(\omega,\vec{x}_\perp)=\sqrt{\frac{4Z^2\alpha}{\omega_\gamma}}\int\frac{d^2\vec{k}_{\gamma\perp}}{(2\pi)^2}\vec{k}_{\gamma\perp}\frac{F_\gamma(\vec{k}_\gamma)}{|\vec{k}_\gamma|^2}e^{i\vec{x}_\perp{\cdot}\vec{k}_{\gamma\perp}},\notag\\
	\vec{k}_\gamma=(\vec{k}_{\gamma\perp},\frac{\omega_\gamma}{\gamma_c}),     \qquad \omega_\gamma=\frac{1}{2}M_{\pi^{+}\pi^{-}}e^{\pm{y}},
    \label{photonflux}
    \end{gather}
where $\vec{x}_{\perp}$ and $\vec{k}_{\gamma\perp}$ are two-dimensional photon position and momentum vectors perpendicular to the beam direction, Z is the charge number of the nuclear, $\alpha$ is the electromagnetic coupling constant, $\gamma_{c}$ is the Lorentz factor of the photon-emitting nucleus, $M_{\pi^{+}\pi^{-}}$ and y are the invariant mass and the rapidity of the pion pair, and $F_{\gamma}(\vec{k}_{\gamma})$ is the nuclear electro-magnetic form factor.

To derive the scattering amplitude $\Gamma_{\gamma A\rightarrow\pi^+ \pi^-A}$ in the coordinate space, we first derive the forward scattering cross section for non-resonant $\gamma A\rightarrow \pi^+\pi^- A$ following~\cite{Pumplin:1970kp}. The Feynman diagram for the corresponding process is shown in Fig.\ref{Fynmandiagram}(b), with variable names as indicated in the figure. 
\begin{figure*}[hbpt]
    \centering
    \includegraphics[width=0.47\linewidth]{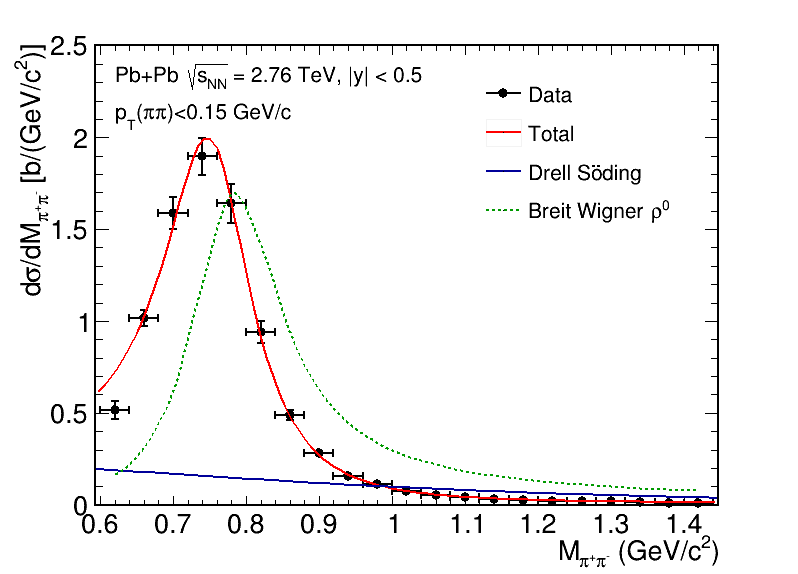}
    \includegraphics[width=0.47\linewidth]{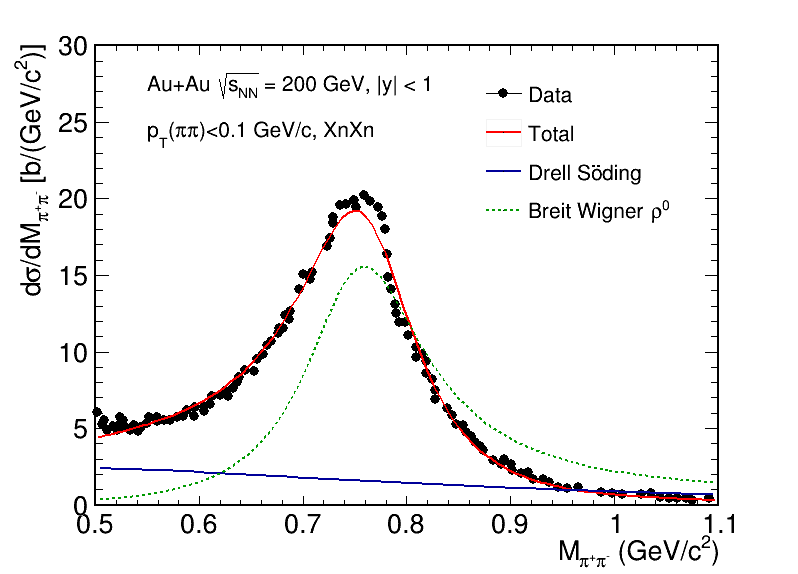}
    \caption{Comparison between the calculated $\pi^+\pi^-$ invariant mass spectrum and experimental data. The left panel is the result for Pb+Pb collisions at 2.76 TeV and the right panel is the result for Au+Au collisions at 200 GeV. The blue solid curve represents the calculated Drell-S${\rm\ddot{o}ding}$ contribution, while the green dashed line indicates the resonant $\rho^0$ contribution. The red curve shows the total contribution, which provides excellent agreement with experimental data (black points).}
    \label{Drell_mass}
\end{figure*}

The forward cross section for $\frac{d\sigma_{\gamma p \rightarrow \pi^{+}\pi^{-} p}}{dtdM_{\pi^{+}\pi^{-}}}|_{t=0}$ can be written as:
\begin{equation}
    \label{crosssection1}
    \frac{d\sigma}{dtdM_{\pi^+\pi^-}}|_{t=0}=\int \frac{q}{256\pi^{4}\left(s-m_{p}^{2}\right)^{2}}\left|M\right|^{2}d\Omega,
\end{equation}
where 
\begin{align}
\mathrm{M} = e \biggl[ 
    &\left(\frac{\epsilon\cdot q_{+}}{k\cdot q_{+}}\right)T_{-} 
     - \left(\frac{\epsilon\cdot q_{-}}{k\cdot q_{-}}\right)T_{+} \notag\\
    &\qquad\qquad+ \frac{\epsilon\cdot (p+p')}{k\cdot (p+p')}(T_{+}-T_{-})
\biggr]*\sqrt{P_{survived}},
\label{Matrixelement}
\end{align}
$\epsilon$ is the photon polarization vector. For Heavy-Ion collisions case, the polarization vector is almost aligned with the impact parameter, $\epsilon$=($\pm$1,0,0;0). $T_{+}$, $T_{-}$ are pion-nucleon elastic scattering amplitudes for initial state pions which are slightly off the mass shell, written as

\begin{gather}
    \label{T-matrix}
    T_{\pm}=2\textit{i}F\sigma_{\pi A},
\end{gather}
$F=\left[\left(q_{\pm}\cdot p'\right)^{2}-m_{\pi}^{2}m_{p}^{2}\right]^{1/2}$ is the invariant flux factor, $\sigma_{\pi A}$ is the total cross section of $\pi A$ scattering which can be calculated using Glauber model~\cite{Miller:2007ri}:
\begin{equation}
    \sigma_{\pi A}=2\int\left(1-\exp\left(-\frac{\sigma_{tot}*T\left(\vec{x}_\perp\right)}{2}\right)\right)d^2\vec{x}_\perp.
\end{equation}
$\sigma_{tot}$ is the total $\pi$--$p$ scattering cross section~\cite{Liu:2023tjr}. The off-shell correction for the $\pi$--$p$ scattering cross section:
\begin{equation}
    F_\pi(Q^2)=\frac{1}{1+\frac{Q^2}{\Lambda^2}},
\end{equation}
is taken into account, where $\Lambda^2=1.5$ $\text{GeV}^2$ is relate to the next resonance from the Regge $\pi$-meson trajectory. 
$P_{survived}$ is the probability that the real pion directly generated from photon fluctuation is not absorbed by the atomic nucleus, which can also be given by the Glauber model.

The relationship between $\Gamma_{\gamma A\rightarrow \pi^+ \pi^- A}\left(M_{\pi^+ \pi^-},\vec{x}_{\perp} \right)$ and the forward cross section $\frac{d\sigma}{dtdM_{\pi^+ \pi^-}}|_{t=0}$ is:
\begin{align}
    \frac{d\sigma}{dtdM_{\pi^+ \pi^-}}|_{t=0}= 
    \frac{1}{4\pi}\left(\int\Gamma\left(M_{\pi^+ \pi^-},\vec{x}_{\perp} \right)d^2\vec{x}_{\perp}\right)^2.
\end{align}
Thus the scattering amplitude can be obtained by combining~\cref{crosssection1,Matrixelement} and differentiating with respect to the coordinates.

The 2D-transverse momentum distribution of the pion pairs from DS process can be obtained by performing a Fourier transformation of the coordinate space amplitude:
\begin{equation} 
\begin{aligned}
    \frac{d^{3}P}{dM_{\pi\pi}dp_{x}dp_{y}} = \bigg| & \frac{1}{2\pi} \int d^{2}x_{\perp} \bigl[ A_{1}(x_{\perp}) \\
    &\qquad\qquad\qquad + A_{2}(x_{\perp}) \bigr] e^{ip_{\perp}\cdot x_{\perp}} \bigg|^{2}.
\end{aligned}
\label{probdis}
\end{equation}
The cross section can be obtained by integrating the impact parameter in  Eq.~\ref{probdis}.

Figure~\ref{Drell_mass} shows that the present framework provides a consistent description of the measured $\pi^+\pi^-$ invariant-mass spectra at both LHC~\cite{ALICE:2015nbw} and RHIC~\cite{STAR:2017enh} energies. The coherent superposition of the resonant $\rho^0$ amplitude and the calculated DS contribution (red curves) reproduces  the experimental distributions (black points) over a wide kinematic range, which has not been previously achieved within a single, unified theoretical framework. The same theoretical setup, including the ultra-peripheral collision selection and mutual Coulomb dissociation conditions employed in the experimental analyses, is applied at both energies. As discussed above, the interference between the DS continuum (blue curves) and the resonant $\rho^0$ amplitude (green curves) induces a characteristic distortion of the invariant-mass spectrum relative to a pure Breit–Wigner shape, providing a practical handle for experimentally isolating the DS contribution.

Having established that the DS contribution is quantitatively under control, we now turn to observables that directly prove the underlying quantum interference and entanglement structure of the produced DS pair. As a direct consequence of the linear polarization of the incident photons and the coherent superposition of the two production sources, the DS pairs exhibit a characteristic second-order azimuthal modulation in the decay angular distribution,
\begin{equation}
\frac{d^2N}{d\cos\theta\,d\phi}
=
\frac{3}{8\pi}\sin^2\theta
\left[1+\cos2(\phi-\Phi)\right],
\end{equation}
where $\Phi$ denotes the angle between the photon polarization vector and the $x$ axis, and $\theta$ and $\phi$ are the polar and azimuthal angles of the pion in the $\pi\pi$ rest frame. This polarization-induced modulation provides a direct and experimentally accessible signature of the interference-induced quantum entanglement discussed above.
\begin{figure}
    \centering    \includegraphics[width=1.0\linewidth]{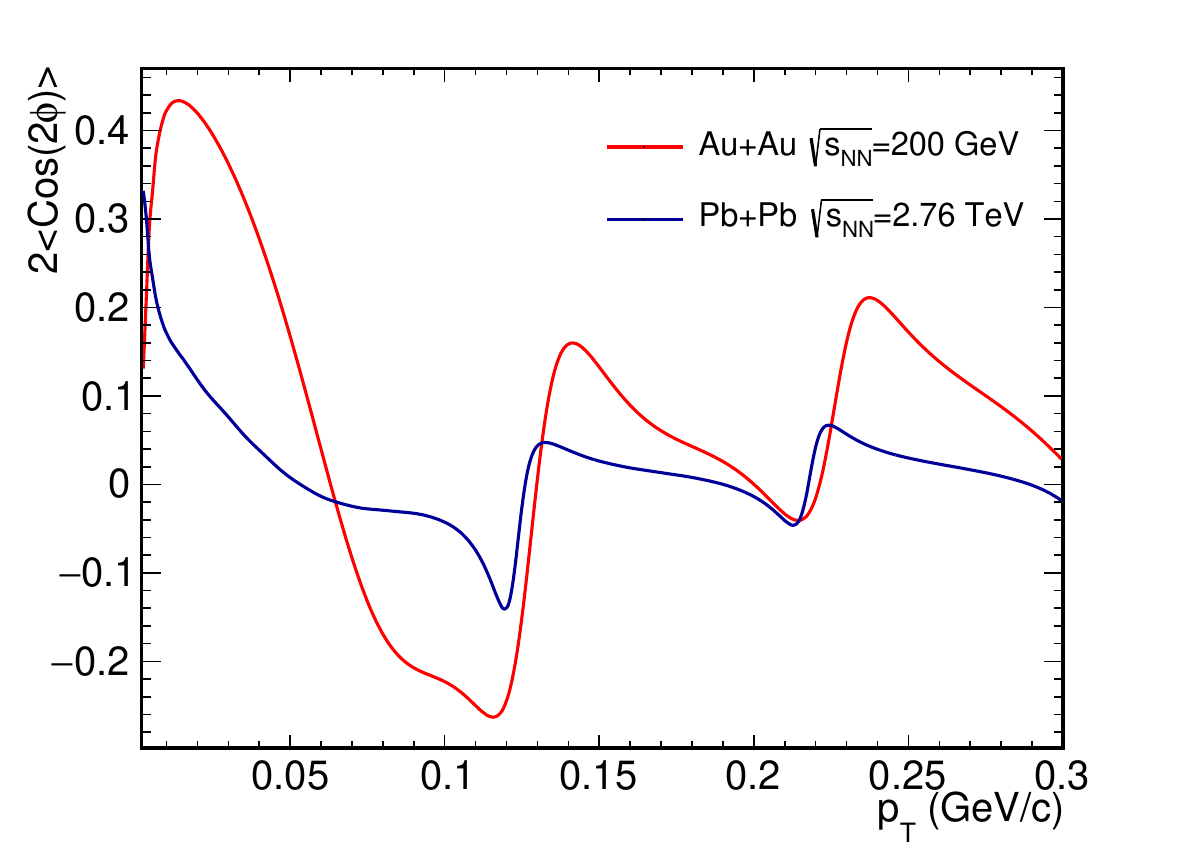}
    \caption{The second-order azimuthal modulation $2\langle \cos 2\phi \rangle$ of DS $\pi^+\pi^-$ pairs as a function of transverse momentum in ultra-peripheral heavy-ion collisions, arising from the interference of linearly polarized photoproduction amplitudes.}
    \label{cos2phi}
\end{figure}


Figure~\ref{cos2phi} shows the calculated $2\langle \cos 2\phi\rangle$ as a function of the transverse momentum of the DS pion pair for RHIC and LHC energies. A pronouced oscillatory modulation is observed, arising from the combined effect of diffractive scattering and quantum interference between the two spatially separated photoproduction sources. In Pb+Pb collisions, the higher collision energy corresponds to ultra-peripheral interactions occurring at larger typical impact parameters, which effectively compresses the interference fringes and shifts the first principal maximum toward smaller transverse momentum. At the same time, the larger spatial extent of the Pb nucleus compared to that of Au corresponds to a wider effective slit in the interference picture, which reduces the contrast between bright and dark fringes. In addition, the increased nuclear size leads to a broader distribtuion of the photon polarization angle $\Phi$, further diluting the polarization-induced azimuthal modulation. As a result, the amplitudes of the maxima and minima in the $2\langle \cos 2\phi\rangle$ modulation for Pb+Pb collisions are smaller than those observed in Au+Au collisions. 

Consequently, smaller systems are in principle more advantageous for extracting the $2\langle \cos 2\phi\rangle$ modulation. However, the reduced nuclear charge and mass numbers in such systems suppress both the photon flux and the $\pi A$ scattering cross section, implying that statistical limitations will need to be carefully considered in future O+O or Ne+Ne collision measurements. More broadly, the ability to tune the two-source geometry and the polarization pattern in UPCs provides a controlled handle to study the quantum-to-classical transition in a relativistic setting. Future measurements that vary the system size and collision energy, and exploit neutron-tagging to constrain the impact-parameter distribution~\cite{Wu:2025dxg}, can map how the azimuthal modulation evolves as which-source information and environmental averaging are gradually introduced. On the theory side, extending the present amplitude-level framework to a density-matrix description will enable quantitative modeling of decoherence, e.g., through phenomenological coherence factors or open-system evolution kernels that suppress interference terms. A systematic comparison between the effectively zero-lifetime DS continuum and finite-lifetime resonant channels (such as $\rho^0$ and $J/\psi$) would then provide a calibrated baseline-to-signal program to isolate genuine decoherence effects from kinematic or acceptance smearing.



In summary, we have presented the first theoretical description of the DS contribution in relativistic heavy-ion collisions and demonstrated that coherent photoproduction in ultra-peripheral collisions naturally generates particle pairs with effectively zero lifetime. In this regime, no intermediate time evolution is available to modify relative quantum phases, allowing production stage quantum coherence to be directly reflected in final-state observables. The coherent superposition of amplitudes from two indistinguishable photon sources, together with the linear polarization of the incident photons, produces a nonseparable two-particle quantum state and imprints a characteristic azimuthal modulation on the momentum distribution of the final-state pion pairs. Owing to the instantaneous nature of the DS process, this modulation is robust against final-state interactions and preserves the underlying quantum interference structure. Our results establish ultra-peripheral heavy-ion collisions as a controlled and experimentally accessible platform for probing interference-induced quantum entanglement in relativistic environments, opening new opportunities for testing quantum coherence in high-energy processes.

\bigskip

This work is supported in part by the National Key Research and Development Program of China under Contract
No. 2022YFA1604900 the National Natural Science Foundation of China (NSFC) under Contract No. 12422510 and 12175223. W.Z. is supported by Anhui Provincial Natural
Science Foundation No. 2208085J23 and Youth Innovation
Promotion Association of Chinese Academy of Sciences. X. Li is supported by National Natural Science Foundation of China (NSFC) under grant No. 124B2103.

\bibliographystyle{unsrtnat}
\bibliography{references}
\end{document}